\documentclass[aps,prx,twocolumn,amsmath,amssymb,superscriptaddress,floatfix]{revtex4-2}
\usepackage{mathtools}
\usepackage{multirow}
\usepackage{xcolor}
\usepackage{bm}
\usepackage{amsfonts,amssymb,amsmath}
\usepackage{graphicx,dcolumn,bm,xcolor,braket,slashed}
\usepackage{times} 
\usepackage{comment}
\usepackage{array}
\usepackage{textcomp}
\usepackage[normalem]{ulem}
\usepackage{dsfont}

\newcommand{\kb}{k_{\mathrm B}}
\newcommand{\cC}{\mathfrak C}

\begin{document}

\title{Topological Transverse Transport without a Gap in Critical Topological Flat Bands}

\author{Chang-geun Oh}
\email{cg.oh.0404@gmail.com}
\affiliation{Department of Applied Physics, The University of Tokyo, Tokyo 113-8656, Japan}
\author{Tomoki Ozawa}
\email{tomoki.ozawa.d8@tohoku.ac.jp}
\affiliation{Advanced Institute for Materials Research (WPI-AIMR), Tohoku University, Sendai 980-8577, Japan}
\affiliation{RIKEN Center for Interdisciplinary Theoretical and Mathematical Sciences (iTHEMS), RIKEN, Wako, Saitama 351-0198, Japan}
\author{Jun-Won Rhim}
\email{jwrhim@ajou.ac.kr}
\affiliation{Department of Physics, Ajou University, Suwon 16499, Republic of Korea}

\begin{abstract}
Quantized Hall transport is traditionally anchored to a bulk spectral gap, which isolates the occupied subspace and exponentially suppresses thermal deviations. Recently discovered critical topological flat bands (CTFBs) challenge this paradigm: an exactly flat band touches a dispersive continuum while retaining a well-defined integer Chern number. However, their finite-temperature transverse transport properties remain entirely unexplored. Here, we develop a low-temperature theory of the intrinsic electrical, thermoelectric, and thermal Hall responses in CTFBs. At fixed particle number, the macroscopic flat-band degeneracy forces a singular Lambert-$W$ drift of the chemical potential, generating an algebraic-logarithmic hierarchy of low-temperature corrections: $T\ln(1/T)$ for electrical Hall, $T[\ln(1/T)]^2$ for thermoelectric Hall, and $T[\ln(1/T)]^3$ for thermal Hall conductivity, replacing the activated thermal protection of a gapped Chern insulator. By contrast, externally pinning the chemical potential to the flat-band energy locks the flat band to half occupation at any nonzero temperature, obstructing the recovery of the fully filled topological ground state as $T\to0^+$. Our results establish that a bulk spectral gap is unnecessary for zero-temperature Hall quantization, but indispensable for its exponential thermal protection.
\end{abstract}

\maketitle

\textit{Introduction.--}
The extreme suppression of kinetic energy in electronic systems provides fertile ground for uncovering correlated and topological quantum phenomena~\cite{LeykamReview,DerzhkoRev,aoki2025flat,checkelsky2024flat}.
In dispersionless or narrow bands, Coulomb interactions dominate over band kinematics, fostering a wide spectrum of exotic states including flat-band ferromagnetism~\cite{Mielke,Tasaki}, fractional Chern insulators~\cite{Tang2011,Sun2011,Neupert2011,Xie2021}, and correlated phases alongside unconventional superconductivity in magic-angle twisted bilayer graphene~\cite{CaoIns,CaoSC}.
Remarkable experimental strides have brought these possibilities into reality, most notably through the observation of fractional quantum anomalous Hall states in twisted MoTe$_2$~\cite{cai2023signatures,zeng2023thermodynamic} and rhombohedral multilayer graphene~\cite{lu2024fractional}.
With kinetic energy quenched, the macroscopic physics is no longer governed by Fermi velocity, but is instead dictated by the internal structure and quantum geometry of the Bloch wave functions---as exemplified by the geometric origin of the superfluid weight in flat-band superconductors~\cite{peotta2015superfluidity,torma2022superconductivity,yu2025quantum}.

In the ideal limit of exact flatness, such bands can be systematically classified according to the regularity and topology of their band projector~\cite{li2026stable,liu2026theory,oh2026classification}.
A trivial, nonsingular flat band admits a globally regular projector~\cite{li2026stable,oh2026classification} and carries no stable topological invariant~\cite{ChenMazaheriSeidelTang2014}.
By contrast, a singular flat band necessarily touches another band at isolated momenta, where its projector becomes direction-dependent and discontinuous~\cite{li2026stable,oh2026classification,rhimkimyang2020,rhimyang2019,RhimYang2021}.
The resulting singularity prevents the flat band from defining an independent Bloch bundle, thereby rendering its individual-band topological invariant ill-defined.
Far from being a mere mathematical pathology, the geometric singularity at these touching points manifests in diverse physical consequences, encompassing bulk-interface correspondence~\cite{oh2022bulk,kim2023general}, anomalous Landau-level spreading under magnetic fields~\cite{rhimkimyang2020,RhimYang2021,long2026interplay, oh2024revisiting}, and a wide range of unconventional transport and response characteristics~\cite{oh2024thermoelectric,oh2026mass,oh2025color,oh2026third,kim2026ultrafast,oh2026magnetic,long2026quantum}.
Bridging these limits, critical topological flat bands (CTFBs) occupy a previously missing intermediate regime characterized by a continuous, albeit nonanalytic, projector~\cite{li2026stable,liu2026theory,oh2026classification,yang2025fractional,yang2025fractional2}.
Although an exactly flat Chern band cannot be both spectrally isolated and generated by strictly local hopping~\cite{ChenMazaheriSeidelTang2014}, a CTFB circumvents this obstruction by touching a dispersive band while preserving projector continuity.
Consequently, the flat band carries a well-defined integer Chern invariant.

The coexistence of well-defined topology and gaplessness separates two roles that are inseparably tied together in a conventional Chern insulator.
There, the bulk gap spectrally isolates the occupied subspace underlying an integer Chern invariant and a quantized Hall conductance~\cite{thouless1982quantized,haldane1988model}, while also providing a finite excitation threshold that exponentially suppresses low-temperature corrections to the quantized response~\cite{Tutschku2020temp}.
In a CTFB, continuity of the flat-band subspace preserves the first property despite the band touching~\cite{li2026stable,liu2026theory,oh2026classification}.
At complete flat-band filling, the resulting well-defined Chern number implies a quantized zero-temperature intrinsic Hall conductivity~\cite{thouless1982quantized}.
The second property, however, is absent because the dispersive continuum touches the occupied flat band without an excitation threshold.
This disparity raises a fundamental question: \emph{how does a topologically quantized but gapless state approach its zero-temperature transport limit, and what replaces the activated thermal protection of a gapped Chern insulator?}

In this Letter, we establish the low-temperature transverse transport theory of CTFBs.
We introduce the cumulative Chern weight, defined as the Berry curvature accumulated over all states below a given energy cutoff, which directly dictates the zero-temperature intrinsic Hall response at the corresponding filling.
Using this quantity, we systematically analyze the Hall, thermoelectric Hall, and thermal Hall responses, contrasting them directly with those of a gapped Chern insulator.
At fixed particle number corresponding to complete flat-band filling, the CTFB recovers the Chern-insulator transport limits at zero temperature $T=0$, while its leading finite-temperature deviations follow the algebraic-logarithmic hierarchy $T[\ln(1/T)]^n$: $n=1$ for the Hall response, $n=2$ for the thermoelectric Hall response, and $n=3$ for the thermal Hall response.
This hierarchy originates from the singular temperature dependence of the chemical potential required by particle-number conservation. By contrast, if the chemical potential is externally pinned to the flat-band energy, the macroscopic degeneracy fixes every flat-band state at half occupation for any nonzero temperature. The resulting grand-canonical state therefore does not approach the completely filled flat-band ground state as $T\to0^+$, and its transport is governed by the half-filled flat-band ensemble.

\begin{figure}[t]
\includegraphics[width=85mm]{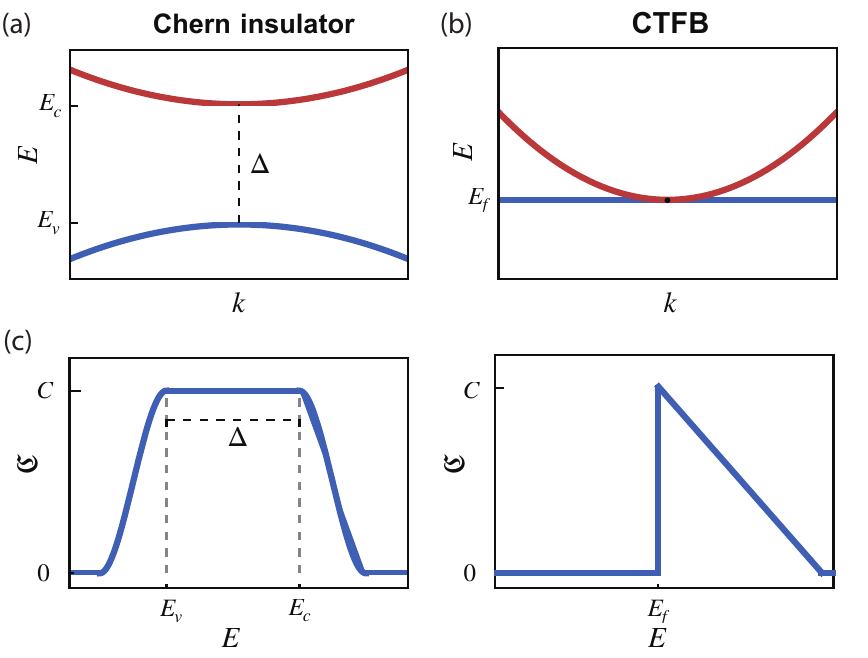} 
\caption{
\textbf{Schematic comparison of a gapped Chern insulator and a critical topological flat band}.
(a)~Band structure of a gapped Chern insulator with a spectral gap $\Delta = E_c - E_v$ separating the valence and conduction bands.
(b)~Band structure of a CTFB system, where an exactly flat band at $E_f$ (blue) touches an upper dispersive band (red) quadratically at a single momentum point.
(c,d)~Cumulative Chern weight $\cC(E)$ for (c) Chern insulator and (d) CTFB system.
Here, $C$ is an integer.}\label{fig1}
\end{figure}

\textit{Setup.--}
We introduce the cumulative Chern weight
\begin{equation}
\cC(E) = \sum_n\frac{1}{2\pi}
\int_{\mathrm{BZ}} d^2k\, \Omega_n(\bm{k})\,
\Theta\!\left(E-\varepsilon_{n\bm{k}}\right),
\label{eq:Ccum}
\end{equation}
where $\Theta(x)$ denotes the Heaviside step function. The Bloch eigenstates satisfy $\hat{H}(\bm{k})|u_{n\bm{k}}\rangle = \varepsilon_{n\bm{k}}|u_{n\bm{k}}\rangle$, and the Berry curvature of band $n$ is defined by $\Omega_n(\bm{k})=
i\left[ \langle \partial_{k_x}u_{n\bm{k}}|\partial_{k_y}u_{n\bm{k}}\rangle-\langle \partial_{k_y}u_{n\bm{k}}|\partial_{k_x}u_{n\bm{k}}\rangle \right]$.
Note that, because we also consider partially filled bands, the
momentum-resolved Berry curvature must be evaluated in a convention that incorporates the physical intracell positions of the orbitals~\cite{oh2026orbital,simon2020contrasting}.
The energy argument $E$ serves as a spectral cutoff that defines the occupied subspace, $\varepsilon_{n\bm{k}} \le E$, thereby parametrizing the zero-temperature band filling.
At zero temperature $T=0$, the cumulative Chern weight is proportional to the intrinsic Hall conductivity as
\begin{equation}
\sigma_{xy}^{(0)}(E) = \frac{e^2}{h}\cC(E),
\label{eq:hallzero}
\end{equation}
where $e$ is the electron's charge, and $h$ is the Planck's constant.

At finite temperature, the intrinsic transverse transport coefficients can be expressed as thermal convolutions of $\cC(E)$~\cite{SM,xiao2006berry,qin2011energy}:
\begin{align}
\sigma_{xy} &= \frac{e^2}{h} \int_{-\infty}^{\infty} dx\,
w(x)\,
\cC(\mu+\kb T x),,
\\
\alpha_{xy} &= -\frac{e\kb}{h}
\int_{-\infty}^{\infty} dx\,x\,w(x)\, \cC(\mu+\kb T x),
\label{eq:alphafinite}\\
\frac{{\kappa}_{xy}}{T} &= \frac{\kb^2}{h}
\int_{-\infty}^{\infty} dx\,x^2w(x)\, \cC(\mu+\kb T x),
\label{eq:kappafinite}
\end{align}
where $w(x)= 1/[4\cosh^2(x/2)]$, $\mu$ is the Fermi level, and $k_B$ is the Boltzmann constant.
Here $\sigma_{xy}$ and $\alpha_{xy}$ denote the intrinsic electrical Hall and thermoelectric Hall conductivities~\cite{xiao2006berry}, while $\kappa_{xy}$ is the thermal Hall conductivity defined at vanishing electric field~\cite{qin2011energy}.

\textit{Gapped Chern insulator.--}
For a conventional gapped Chern insulator, when the zero-temperature Fermi energy $E$ lies inside the bulk gap, the cumulative Chern weight remains fixed at an integer value $C$ throughout a finite energy interval, as illustrated in Fig.~\ref{fig1}:
\begin{equation}
\cC(E)=C\in \mathbb{Z},
\qquad
E_v<E<E_c,
\end{equation}
where $E_v$ and $E_c$ denote the valence- and conduction-band edges, respectively. 
Consequently, every derivative of $\cC(E)$ vanishes in a finite neighborhood of the chemical potential. All algebraic terms in the Sommerfeld expansion therefore vanish. 
In the zero-temperature limit, the transverse transport coefficients approach~\cite{qin2011energy,xiao2006berry}
\begin{equation}
\sigma_{xy}^{(0)} = C\frac{e^2}{h},
\quad
\alpha^{(0)}_{xy}=0,
\quad
\lim_{T\rightarrow0}
\frac{{\kappa}_{xy}(T)}{T}
=
C\frac{\pi^2\kb^2}{3h}.
\label{eq:ZeroT}
\end{equation}
The remaining low-temperature corrections arise from thermally excited carriers reaching the band edges. 
Defining the energy gap $\Delta=E_c-E_v$
we obtain the generic activated behavior as~\cite{SM}
\begin{align}
\delta\sigma_{xy} &\equiv
\sigma_{xy}(T)-C\frac{e^2}{h}\sim
Te^{-\Delta/(2\kb T)}, \nonumber\\
\delta\alpha_{xy}
&\equiv \alpha_{xy}(T)-\alpha_{xy}^{(0)}
\sim \Delta e^{-\Delta/(2\kb T)},
\nonumber\\
\delta\!\left(\frac{\kappa_{xy}}{T}\right)
&\equiv
\frac{\kappa_{xy}(T)}{T} - C\frac{\pi^2\kb^2}{3h}
\sim
\frac{\Delta^2}{T}e^{-\Delta/(2\kb T)}.
\label{eq:activated}
\end{align}
Thus the bulk gap imposes a finite energy cost for accessing states outside the Chern plateau, exponentially suppressing all deviations from the zero-temperature transverse responses.
Note that in Eq.~\eqref{eq:activated}, it is considered either
fixed particle number at the insulating filling or a chemical
potential pinned at the midgap energy.

\textit{CTFB.--}
In contrast to a gapped Chern insulator, a CTFB has no finite energy interval over which the cumulative Chern weight remains quantized~\cite{li2026stable,liu2026theory,oh2026classification}.
Instead, the quantized plateau collapses to the single critical energy at which the flat band is completely filled, as illustrated in Fig.~\ref{fig1}. 
Immediately above $E_f$, the gapless dispersive continuum begins to contribute to $\cC(E)$.
Consequently, the finite-temperature dependence of the transverse responses is qualitatively different from that of a gapped Chern insulator.

To demonstrate this distinction, we fix the particle density such that the flat band is completely filled at zero temperature, corresponding to $E=E_f^+$, where $E_f$ denotes the flat-band energy. 
Hereafter, we choose the flat-band energy as the zero of energy,
$E_f=0$.
In the zero-temperature limit, the transverse transport coefficients approach the same quantized values as those of the gapped Chern insulator in Eq.~\eqref{eq:ZeroT}. 
At any finite temperature, however, particle-number conservation requires thermally excited holes in the flat band are compensated by electrons in the touching dispersive band.
The density of thermally generated holes in the flat band is $n_h = 1/[1+e^{\mu/(\kb T)}]$, whereas the density of electrons excited into the quadratically touching dispersive band with low-energy density of states $D_0$ is $n_e = D_0\int_0^\infty \mathrm{d} E\, \frac{1}{e^{(E-\mu)/(\kb T)}+1}=D_0\kb T\ln[1+e^{\mu/(\kb T)}]$.
Defining $Y \equiv \ln[1+e^{\mu/(\kb T)}]$, the fixed-filling condition $n_h = n_e$ simplifies exactly to $e^{-Y} = D_0\kb T\,Y$, yielding
\begin{equation}
Y(T) = W_0\!\left(\frac{1}{D_0\kb T}\right),
\label{eq:LambertY}
\end{equation}
where $W_0$ denotes the principal branch of the Lambert $W$ function.
The chemical potential is therefore
\begin{equation}
\mu(T) = \kb T\ln\!\left(e^Y-1\right) \simeq \kb T\,Y,
\label{eq:muLambert}
\end{equation}
which holds asymptotically for $T\to 0$.
The excited carrier density is correspondingly $n_h = n_e = D_0\kb T\,Y$, vanishing as $T\ln[1/(D_0\kb T)]$ despite the logarithmic divergence of $Y\simeq\ln[1/(D_0\kb T)]$ as $T\to 0$.

To characterize the Chern weight carried by the touching dispersive band,
we expand the cumulative Chern weight above the flat-band as
\begin{equation}
\cC(E) =
C+D_0\widetilde{\Omega}_0E
+\mathcal O\!\left[E^2\right],
\label{eq:ChernWeightExpansion}
\end{equation}
where
\begin{equation}
\widetilde{\Omega}_0
\equiv
\frac{1}{D_0}
\left.
\frac{d\cC(E)}{dE}
\right|_{E=0^+}.
\label{eq:OmegaTilde}
\end{equation}
Here $\widetilde{\Omega}_0$ is dimensionless and measures the Chern weight carried per state for the touching dispersive band.

At $T=0$, complete filling of the flat band guarantees that the transverse transport coefficients are identical to those of the gapped Chern insulator in Eq.~\eqref{eq:ZeroT}.
At $T>0$, the thermally generated flat-band holes and dispersive electrons give the low-temperature corrections~\cite{SM}
\begin{align}
\frac{\delta\sigma_{xy}}{e^2/h}
&= D_0\kb T(\widetilde{\Omega}_0-C)Y
+\cdots,
\nonumber\\
\frac{\delta\alpha_{xy}}{e\kb/h}
&=-D_0\kb T\left[CY(Y+1) +\frac{\pi^2}{3}\widetilde{\Omega}_0 \right]
+\cdots,
\nonumber\\
\frac{\delta\!\left({\kappa}_{xy}/T\right)}{\kb^2/h}
&=
D_0\kb T\,Y\left[\frac{\pi^2}{3}\widetilde{\Omega}_0
-C(Y^2+2Y+2)\right]
+\cdots.
\label{eq:CTFBLambert}
\end{align}
The two contributions to the Hall correction have a simple interpretation: each thermally generated hole removes the Chern weight $C$ of the filled flat band, whereas each electron excited into the touching dispersive band carries the local Chern weight
$\widetilde{\Omega}_0$. Hence the net Hall correction is controlled by $\widetilde{\Omega}_0-C$.
For $T\to 0$, these expressions make the asymptotic hierarchy transparent as
\begin{align}
\delta\sigma_{xy}
&\sim
D_0\kb T \ln\!\left(\frac{1}{D_0\kb T}\right),
\nonumber\\
\delta\alpha_{xy}
&\sim
D_0\kb T \left[ \ln\!\left(\frac{1}{D_0\kb T}\right) \right]^2,
\nonumber\\
\delta\!\left(\frac{{\kappa}_{xy}}{T}\right)
&\sim
D_0\kb T \left[ \ln\!\left(\frac{1}{D_0\kb T}\right) \right]^3.
\label{eq:CTFBHierarchy}
\end{align}
Thus, although a CTFB and a gapped Chern insulator share the same zero-temperature topological responses, they approach the limits through fundamentally different thermal routes: exponentially activated corrections in the gapped insulator versus algebraic-logarithmic corrections in the CTFB.

\begin{figure*}[t]
\includegraphics[width=170mm]{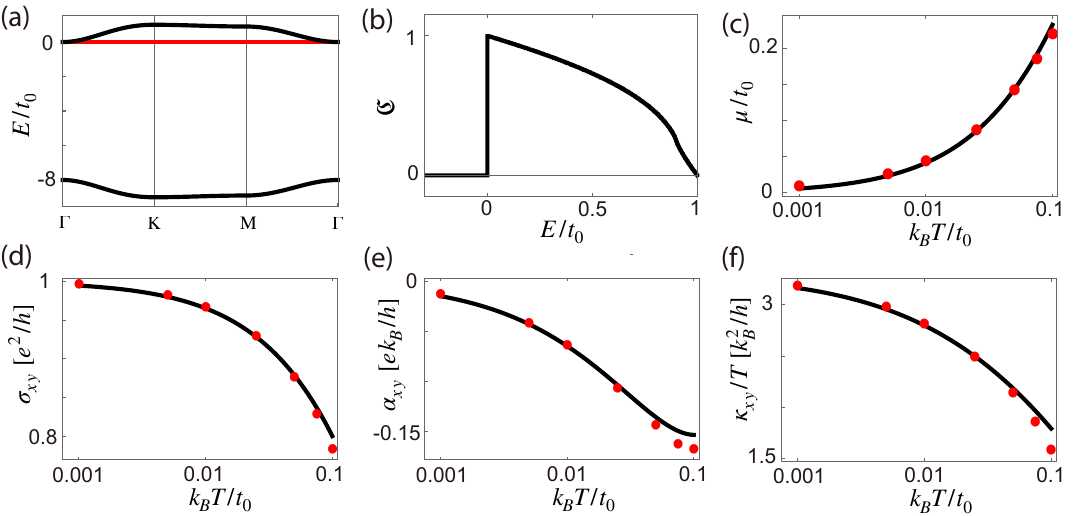} 
\caption{
\textbf{Lattice model verification of the critical topology flat-band continuum theory.}
(a) Band structure of the fluxed dice-lattice model in Eq.~(\ref{eq:diceHam}).
(b) Cumulative Chern weight $\cC(E)$.
(c)--(f) The temperature $T$ dependence of the fixed-filling chemical potential $\mu$, electric Hall $\sigma_{xy}$, thermoelectric Hall $ \alpha_{xy}$, and thermal Hall $ (\kappa_{xy}/T)$ coefficients.
Red circles denote the results from lattice calculations, while
solid curves are the low-energy continuum results in Eqs.~(\ref{eq:muLambert}) and (\ref{eq:CTFBLambert}) with
$D_0$ and $\widetilde{\Omega}_0$ determined from the lattice model. See SM for the derivation detail.
}\label{fig2}
\end{figure*}

\textit{Lattice model verification.--}
To verify the continuum predictions in a microscopic setting, we consider the $2\pi$-fluxed dice lattice model of Ref.~\cite{yang2025fractional2,yang2025fractional}.
The Bloch Hamiltonian is given by
\begin{equation}
H(\bm{k})=-
\begin{pmatrix}
E_A & f(\bm{k}) & g(\bm{k})\\
f^*(\bm{k}) & 0 & 0\\
g^*(\bm{k}) & 0 & 0
\end{pmatrix},
\label{eq:diceHam}
\end{equation}
with $f(\bm{k}) = -t_0 \sum_{j=1}^{3} c_j e^{-i\bm{k}\cdot\bm{d}_j}$, $ g(\bm{k}) = t_0 \sum_{j=1}^{3}
c_j e^{i\bm{k}\cdot\bm{d}_j}$,
where $(c_1,c_2,c_3) =
(1,\omega,\omega^*)$,
$\omega=e^{2\pi i/3}$,
$\bm d_1=(1,0)$,
$\bm d_2=
\left(-\frac12,\frac{\sqrt3}{2}\right)$,
and
$\bm d_3=
\left(-\frac12,-\frac{\sqrt3}{2}\right)$.
We take the bond length as the unit of length and set $E_A = 8t_0$ (with $t_0 > 0$).
Because $\sum_{j=1}^3 c_j = 0$, both $f(\bm{k})$ and $g(\bm{k})$ vanish at $\bm{k}=\bm{0}$, ensuring that the flat band at $E_f=0$ touches the upper dispersive band quadratically at the $\Gamma$ point.

As shown in Fig.~\ref{fig2}(a,b), the resulting band structure features an isolated lowest band that is topologically trivial, separated by a finite gap from an exactly flat middle band at $E_f=0$ with Chern number $C=1$.
The flat band touches the upper dispersive band quadratically at the $\Gamma$ point.
Consequently, the cumulative Chern weight $\cC(E)$ exhibits a discrete unit jump at $E_f$ [Fig.~\ref{fig2}(b)], but immediately departs from unity for $E>E_f$ as states in the touching dispersive band enter the continuum.
The quantized topological response is therefore supported strictly at the single critical energy $E=E_f$.

We fix the filling to two electrons per unit cell— corresponding to completely filled lowest and flat bands at $T=0$—and determine the chemical potential $\mu(T)$ self-consistently by solving the full lattice number equation~\cite{SM}.
Figures~\ref{fig2}(c)--\ref{fig2}(f) compare the self-consistent lattice results (red circles) with the analytical predictions of Eqs.~(\ref{eq:muLambert}) and (\ref{eq:CTFBLambert}) (solid curves), evaluated using $D_0$ and $\widetilde{\Omega}_0$ extracted directly from the lattice band structure without any adjustable parameters.
As the temperature is lowered, the continuum theory quantitatively captures the Lambert-$W$ dependence of $\mu(T)$ and all three transverse transport coefficients.

\textit{Fixed chemical potential.--}
The fixed-filling condition ($n_h=n_e$) is essential both to recover the quantized zero-temperature responses in Eq.~\eqref{eq:ZeroT} and to generate the low-temperature corrections in Eqs.~\eqref{eq:CTFBLambert} and \eqref{eq:CTFBHierarchy}.
For comparison, consider the grand-canonical setting where the chemical potential is pinned at the flat-band energy, $\mu=E_f=0$.
Because all states in the flat band lie precisely at the chemical potential, their Fermi occupation is strictly $f(0)=1/2$ at any finite temperature.
The flat band therefore remains half filled as $T\to0^+$, in sharp contrast to the fixed-filling case where particle-number conservation drives $\mu(T)$ via Eq.~\eqref{eq:muLambert} and forces the flat-band occupation toward unity.
Consequently, the fixed-$\mu$ limit does not approach the fully occupied Chern-band state.
The transverse responses for $\mu=E_f$ are given as~\cite{SM}
\begin{align}
\frac{\sigma_{xy}}{e^2/h}
&= \frac{C}{2} +D_0\widetilde{\Omega}_0\,\kb T\ln 2
+\cdots,
\nonumber\\
\frac{\alpha_{xy}}{e\kb/h}
&= -C\ln 2
- \frac{\pi^2}{6} D_0\widetilde{\Omega}_0\,\kb T
+\cdots,
\nonumber\\
\frac{\kappa_{xy}/T}{\kb^2/h}
&=C\frac{\pi^2}{6}
+\frac{9}{2}\zeta(3)D_0\widetilde{\Omega}_0\,\kb T
+\cdots.
\label{eq:fixedmu}
\end{align}
Hence, the zero-temperature limits are
\begin{equation}
\lim_{T\to0^+}
\left( \frac{\sigma_{xy}}{e^2/h}, \frac{\alpha_{xy}}{e\kb/h},
\frac{\kappa_{xy}/T}{\kb^2/h} \right)
=
\left( \frac{C}{2}, -C\ln 2,
\frac{\pi^2C}{6} \right),
\label{eq:fixedmuLimit}
\end{equation}
rather than the Chern-insulator values $(C,0,\pi^2C/3)$.
The factor of $1/2$ in the electrical and thermal Hall conductivities directly reflects the half occupation of the flat band.
Meanwhile, the finite zero-temperature limit of $\alpha_{xy}$ arises because each half-occupied flat-band state retains a residual entropy of $\kb\ln 2$, with the thermoelectric Hall response precisely measuring this entropy weighted by the Berry curvature.
These behaviors have no counterpart in a conventional gapped Chern insulator, where pinning $\mu$ inside the bulk gap intersects no states and still yields a completely filled valence band as $T\to0$. In a CTFB, by contrast, pinning $\mu=E_f$ places the chemical potential exactly at the energy of a macroscopically degenerate manifold.
Furthermore, the thermal corrections from the touching continuum in Eq.~\eqref{eq:fixedmu} are purely analytic and linear in $T$, exhibiting no logarithmic enhancement.
Fixed filling is therefore doubly indispensable: it drives the system into the fully occupied topological ground state as $T\to 0$, and it dictates the $T[\ln(1/T)]^n$ scaling that governs the approach to that limit.

\begin{figure}[t]
\includegraphics[width=85mm]{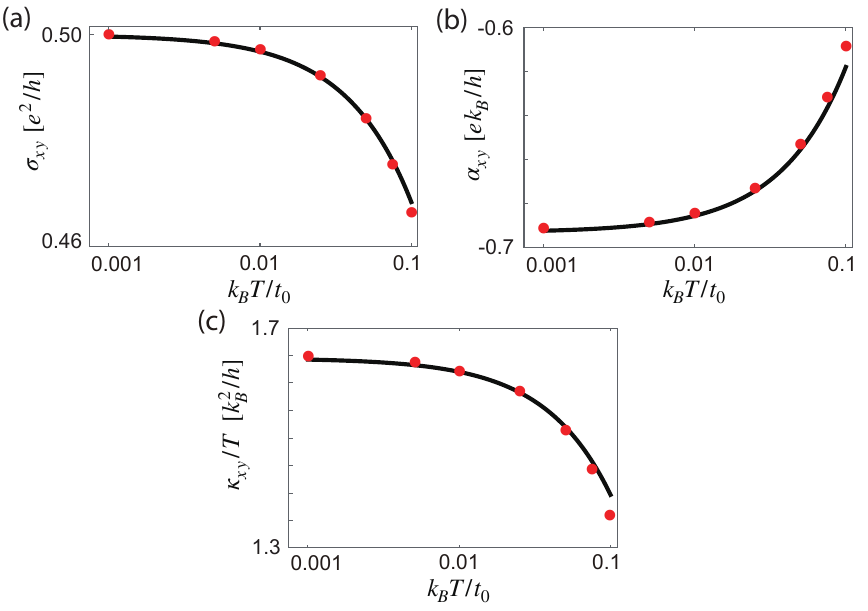} 
\caption{\textbf{Transverse responses at fixed chemical potential.}
Temperature dependence of (a) the electrical Hall conductivity $\sigma_{xy}$, (b) the thermoelectric Hall conductivity $\alpha_{xy}$, and (c) the thermal Hall coefficient $\kappa_{xy}/T$ for the fluxed dice-lattice model at $\mu=E_f=0$. Red circles denote full-lattice calculations, while solid curves show the low-energy continuum results in Eq.~\eqref{eq:fixedmu}, with $D_0$ and $\widetilde{\Omega}_0$ determined from the lattice model~\cite{SM}.}\label{fig3}
\end{figure}

We verify these grand-canonical predictions in the fluxed dice-lattice model of Eq.~\eqref{eq:diceHam} with the chemical potential pinned at $\mu=E_f=0$.
Figure~\ref{fig3} compares the full-lattice calculations (red circles) with the low-energy continuum expressions in Eq.~\eqref{eq:fixedmu} (solid curves).
As $T\to0^+$, the three transverse responses converge to $\sigma_{xy}/(e^2/h)=1/2$, $\alpha_{xy}/(e\kb/h)=-\ln 2$, and $(\kappa_{xy}/T)/(\kb^2/h)=\pi^2/6$ (corresponding to $C=1$), in quantitative agreement with Eq.~\eqref{eq:fixedmuLimit}.
Furthermore, the continuum theory accurately reproduces the leading linear-in-$T$ slopes, corroborating that pinning $\mu$ to the flat band eliminates the logarithmic hierarchy characteristic of fixed filling.
Deviations at higher temperatures stem naturally from $\mathcal{O}(T^2)$ thermal corrections and band dispersion corrections beyond the quadratic low-energy regime.

\textit{Discussion.--}
Our results distinguish zero-temperature quantization from finite-temperature topological protection. A bulk spectral gap is not strictly required for a quantized Hall response at zero-temperature filling, but it is essential for its exponential thermal stability. 
In a conventional Chern insulator, the quantized cumulative Chern weight extends over a finite energy interval, and carriers must be thermally excited across a nonzero gap before the response can deviate from its quantized limit. 
In a CTFB, the dispersive continuum touches the flat band with no excitation threshold, collapsing the quantized plateau to an isolated critical point.
The resulting finite-temperature corrections are therefore algebraic-logarithmic rather than exponentially activated.

Importantly, the logarithmic structure originates from exact flatness,
not from gaplessness alone.
At fixed particle number, the macroscopic degeneracy of the flat band forces a logarithmically enhanced low-temperature dependence of the chemical-potential, $\mu(T)\sim \kb T\ln(1/T)$, which generates the hierarchy of logarithmic enhancements.
If the flat band is instead replaced by a generic dispersive band while the gapless topological touching is retained, the chemical potential shifts only linearly from the touching energy, $\mu(T)\propto T$, and the leading corrections to $\sigma_{xy}$, $\alpha_{xy}$, and $\kappa_{xy}/T$ are generically linear in $T$~\cite{SM}.
Thus, gaplessness removes the exponential thermal protection, whereas exact flatness produces the singular fixed-filling thermodynamics responsible for the hierarchy in Eq.~\eqref{eq:CTFBHierarchy}.

A recent microscopic construction continuously connects the CTFB point to a family of singular flat bands~\cite{yang2025fractional2}.
Although the Hamiltonian varies smoothly along this deformation, the normalized flat-band Bloch states lose continuity at the touching point, and the individual flat-band Chern invariant ceases to be well defined.
Consequently, exact zero-temperature Hall quantization is forfeited, and the transverse response is governed instead by nonquantized Berry flux and the wave-function geometry near the singular touching.
Tracing the temperature evolution of the transverse responses across this crossover therefore offers a promising route to understand how topological quantization gives way to the unconventional transport physics of singular band touchings.

In electronic realizations, interactions, residual flat-band dispersion $W_f$, and disorder $\Gamma$ introduce infrared scales that eventually cut off the critical behavior~\cite{LeykamReview}. 
An interaction-induced gap restores activated transport at the lowest temperatures, while the CTFB scaling governs the intermediate quantum critical window $\max(\Delta_{\mathrm{int}}, W_f, \Gamma) \ll \kb T \ll \Lambda$, below the upper cutoff $\Lambda$ of the quadratic touching. The quantitative agreement with the fluxed dice-lattice model confirms that this regime is not an artifact of the continuum approximation. Simultaneous anomalous Hall, thermoelectric Hall, and thermal Hall measurements would therefore provide a stringent test of the predicted responses in Eq.~(\ref{eq:CTFBLambert}), establishing the exact thermodynamic consequence of gaplessness in topological matter.

\begin{acknowledgments}
The authors thank W. Yang and D. Zhai for useful discussions.
C.Oh acknowledge the support by Japan Society for the Promotion of Science (JSPS), KAKENHI Grant No. JP25KF0186.
J.W.R. was supported by the National Research Foundation of Korea (NRF) Grant funded by the Korean government (MSIT) (grant no. RS-2023-NR068116) and the Ministry of Education (grant no. RS-2023-00285390).
T.O. acknowledges the support by JSPS KAKENHI Grant No. JP24K00548, JST PRESTO Grant No. JPMJPR2353, JST CREST Grant No. JPMJCR26XA, and Frontier Research in Duo (FRiD) grant from Tohoku University.
\end{acknowledgments}

\bibliography{ref}

\end{document}